\documentclass[12pt]{book}

\usepackage[dvips]{graphicx,color}
\usepackage{makeidx,observe,cosmology}
\makeauthorindex
\makeindex

\BookTitle{New Trends in Theoretical and Observational Cosmology}
\CopyRight{\copyright 2001 by Universal Academy Press, Inc.}

\begin{document}

\BookTitle{\itshape New Trends in Theoretical and Observational Cosmology}
\CopyRight{\copyright 2001 by Universal Academy Press, Inc.}
%\tableofcontents
\pagenumbering{arabic}

\chapter{%   %%%%%%%%% <===== TITLE of the contribution
%%%%%%%%%%% The first letter of each word should be captital letter.
How Can We Study the Local Universe with CMB Photons?}

\author{%
Asantha COORAY\\%%% <== First author and second author. This case, same affiliation.
{\it California Institute of Technology, Pasadena, California 91125, USA}}
%\\
%Kaoru Kayashima\\%%%%%% <== Third author. This case, different affiliation.
%{\it Universal Academy Press, Inc., 6-16-2 Hongo, Bunkyo-ku, Tokyo 113-0033, Japan}}
%
% Please note:
% One \AuthorContents{} is necessary
% for EACH CONTRIBUTION, for the contents page and
% One \AuthorIndex{} is necessary
% for EACH AUTHOR, for the index.
%
\AuthorContents{A.\ Cooray} %%%%%%% <=== It is the data for CONTENTS. Please enter all author's name that should be inithialized.

\AuthorIndex{Cooray}{A.}%%%%%%% <=== It is the data for AUTHOR INDEX. Please enter a author's name that should be inithialized.

\section*{Abstract}

We show how observations of temperature fluctuations in cosmic microwave background (CMB) can be used to extract information related to the large scale structure, including
dark matter distribution, pressure and halo velocities involving the line of sight, transverse and rotational components. The frequency spectrum can be used to separate the SZ effect
from thermal fluctuations. The measurement of higher order correlations in CMB can be used to extract effects involving lensing and kinetic SZ effects. A high signal-to-noise arcminute scale CMB experiment can be used to construct the weak lensing convergence, which is useful given that the background source, CMB anisotropies at the last scattering surface, is well understood. 

\section{Introduction}

The cosmic microwave background (CMB) is now well known as a probe of the early universe. The temperature fluctuations in the CMB, especially the acoustic peaks in the angular power spectrum of CMB anisotropies, capture the physics of primordial baryon-photon fluid undergoing oscillations in the potential wells of dark matter \cite{Huetal97}.
 In transit to us, CMB photons also encounter the large scale structure that defines the local universe; thus, several aspects of photon properties, such as the frequency or direction of propagation, are affected.
In the reionized epoch, variations are imprinted when photons are  Compton-scattered via electrons, moving with respect to the CMB.

These gravitational and scattering effects, though some times insignificant compared to primary fluctuations, leave certain imprints in the anisotropy structure while inducing higher order correlations. These signatures can then be used to extract the properties of large scale structure that led to changes in the CMB temperature. 
Here, we will summarize what these signatures are, and, how they can be used to study the local universe with CMB photons.

\begin{figure}[t]
 \begin{center}
    \includegraphics[height=13pc,angle=-90]{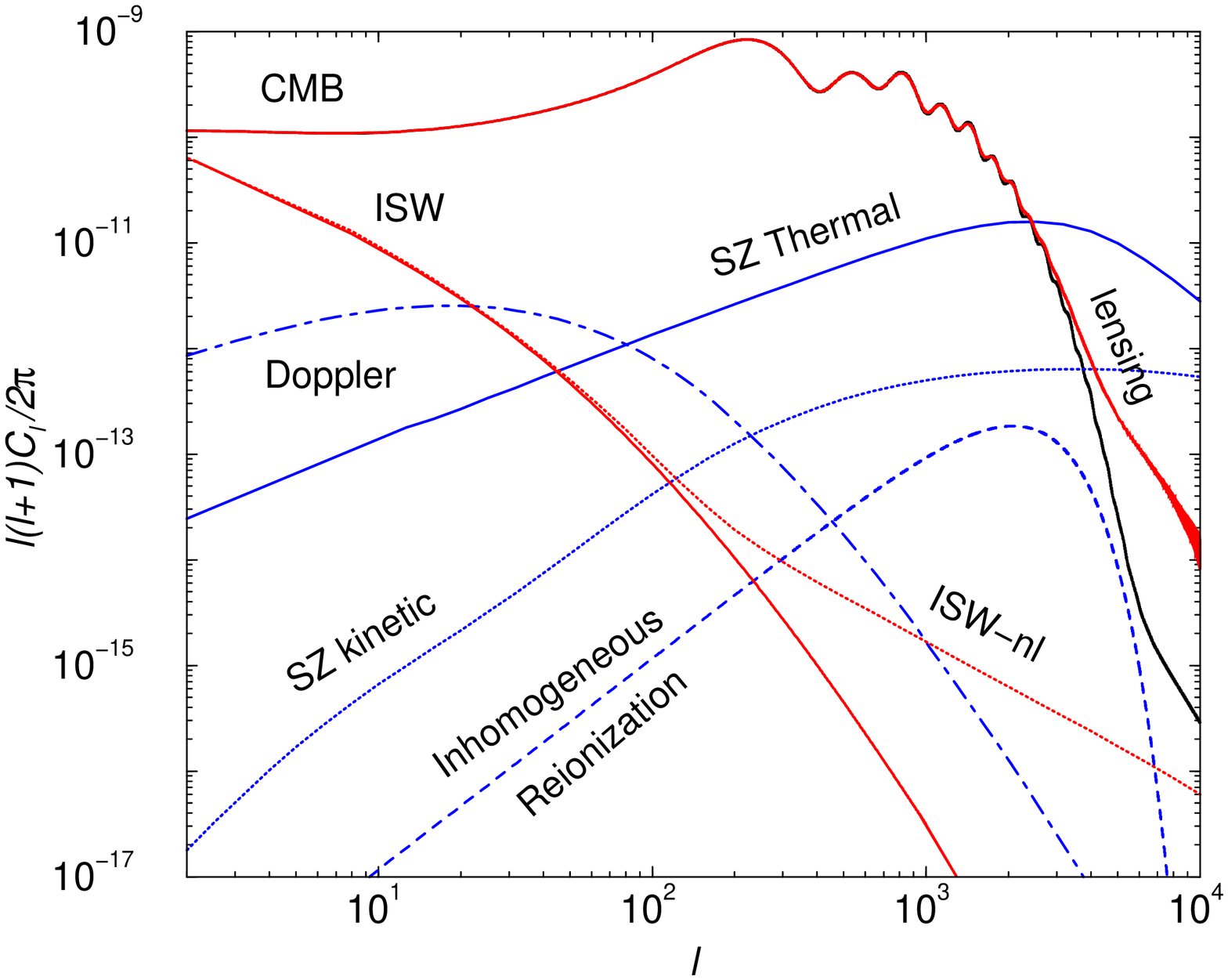}\includegraphics[height=13pc,angle=-90]{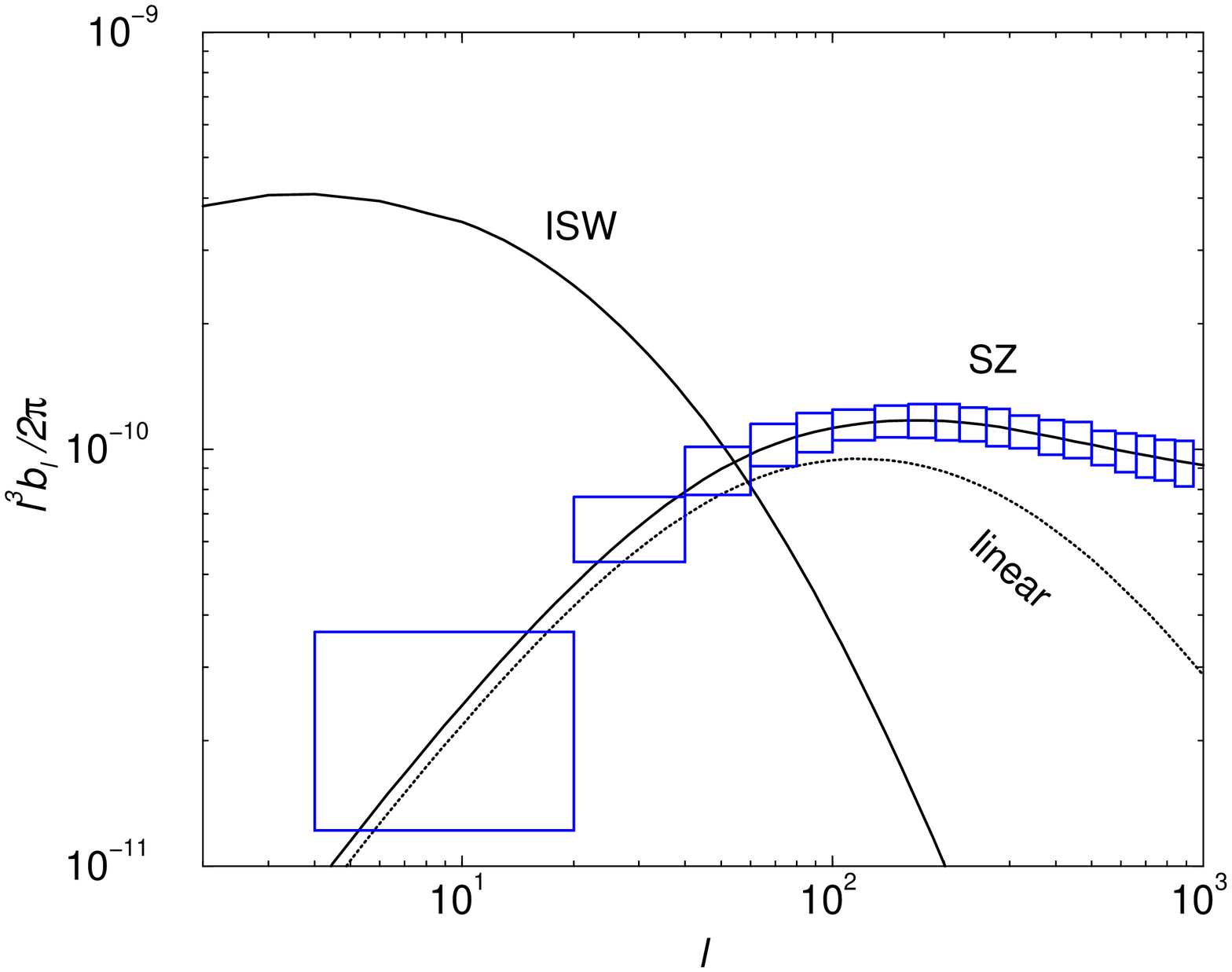}\includegraphics[height=13pc,angle=-90]{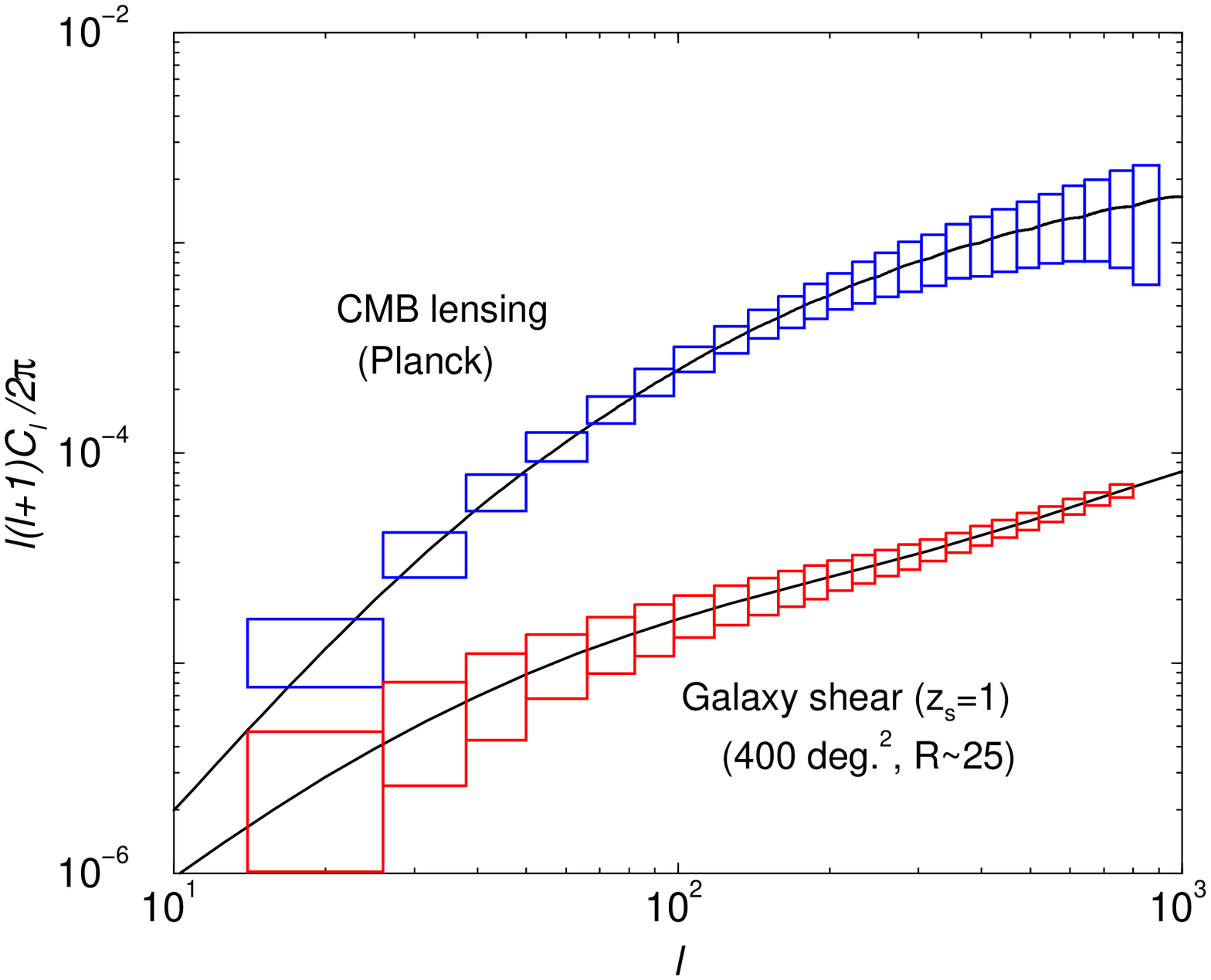}
%%%%% ``13pc'' is just the example.
  \end{center}
\caption{{\it Left}: Power spectrum for the temperature
anisotropies in the fiducial $\Lambda$CDM model
with $\tau=0.1$. The curves show the local universe contributions to CMB due to 
gravity (ISW and lensing) and scattering (Doppler, SZ effects, patchy reionization). 
{\it Middle}: The lensing-secondary correlations and error bars show how well lensing-SZ correlation can be measured with Planck. {\it Left}: CMB as a weak lensing experiment and error bars show the reconstruction of convergence with Planck.}
\label{fig:cl}
\end{figure}

\section{Local Universe Contributions to CMB}

{\it Integrated Sachs-Wolfe Effect:}  The differential redshift effect from photons climbing in and
out of a time-varying gravitational potential along the line of
sight is called the integrated Sachs-Wolfe (ISW; \cite{SacWol67}) effect. 
The ISW effect is important for low matter density universes
$\Omega_m < 1$, where the gravitational potentials decay at low redshift,
and contributes anisotropies on and above the scale of the horizon at the
time of decay. The non-linear contribution to the ISW effect comes from large scale structure 
momentum density field. In \cite{Coo01}, we presented a model for the non-linear contribution;
one aspect of this non-linear effect is that it results from the transverse velocity of halos
across the line of sight and results in a dipolar temperature change aligned with the direction of motion. This contribution, however, 
produces fluctuations which are of order 0.5 $\mu$K when transverse velocities are of order 100 km sec$^{-1}$. The possibility to measure transverse velocities challenges improvements in the experimental front to reach such low temperatures.

{\it Weak Gravitational Lensing:} Gravitational lensing of the photons by the intervening large-scale structure both redistributes power in multipole space and enhances it due power
in the density perturbations.  The most effective structures
for lensing lie half way between the surface of recombination and
the observer in comoving angular diameter distance.  In the
fiducial $\Lambda$CDM cosmology, this is at $z \sim 3.3$,
but the growth of structure skews this to
somewhat lower redshifts.  In general, the efficiency of lensing is
described by a broad bell shaped function between the source and the
observer, and thus, correlates well with a large number of tracers
of the large scale structure  from low to high redshifts.

Since the lensing effect involves the angular gradient of CMB photons and leaves the surface brightness unaffected, its  signatures are at the second order in temperature. A quadratic statistic can be devised to probe the gradient structure of CMB temperature and to extract the projected line of sight dark matter density field. This can be best achieved with the divergence of the temperature weighted gradient statistic of \cite{Hu01} 
and we show errors for a construction of convergence in
figure~1.
Since lensing  has a distinct non-linear mode coupling behavior,
it can produce a non-Gaussian signature in CMB data. The idea behind here is that the potentials which lensed CMB also contribute to first order temperature fluctuations from effects such as
ISW.  These correlations are discussed in detail in \cite{GolSpe99}. 

{\it Sunyaev-Zel'dovich Thermal Effect:} At small angular scales, probably, the best  known effect is the thermal Sunyaev-Zel'dovich 
(SZ; \cite{SunZel80}) effect.
The SZ effect arises from the  inverse-Compton scattering of CMB photons by hot electrons
along the line of sight. This effect has now been directly imaged
toward massive galaxy clusters, where temperature of the scattering medium  can reach as high as 10 keV producing temperature changes in the CMB of order 1 mK.
Note that the effect is proportional to the temperature weighted electron density, or
pressure in a cluster. Using a distribution of halos with gas in hydrostatic equilibrium, we can calculate the pressure clustering and project it along the line of sight to obtain the resulting angular power spectrum of temperature fluctuations \cite{Coo01a}.

 Note that the SZ effect also bears a spectral signature that differs from the other effects.
The upscattering of photons moves them from low to high frequencies, with no effect at
a frequency of $\sim$ 217 GHz. This leads to decrements at low frequencies and increments at high frequencies. An experiment such as Planck with sensitivity
beyond the peak of the spectrum can separate out the SZ contribution
based on the spectral signature \cite{Cooetal00}. This is important since statistics related to the SZ effect can be studied separately uncontaminated by primary anisotropies and confusing foregrounds. For example, in \cite{Coo01b}, we introduced the CMB$^2$-SZ power spectrum as a probe of the lensing-SZ correlation which involves the CMB and frequency cleaned SZ maps.
This statistic can be used to directly estimate how pressure correlates with dark matter;
in figure~\ref{fig:cl}, we show the expected errors for the Planck mission.

{\it Linear Doppler Effect:} 
The bulk flow of the electrons that scatter the CMB photons leads to
a Doppler effect.  Its effect on the power spectrum  peaks around the
horizon at the scattering event projected on the sky
today (see Fig.~\ref{fig:cl}).
On scales smaller than the horizon at scattering, the contributions are
significantly canceled as photons scatter against the crests and troughs of
the perturbation.  As a result, the Doppler effect is moderately sensitive
to how rapidly the universe reionizes since contributions from a sharp
surface of reionization do not cancel.   

{\it Kinetic Sunyaev-Zel'dovich Effect:} The Ostriker-Vishniac/kinetic SZ (kSZ) effect arises from the second-order modulation of the Doppler effect by density fluctuations
\cite{OstVis86}.   Due to the density weighting, 
kSZ effect peaks at small scales  and avoids cancellations
associated with the linear effect; the linear effect can also be modulated via
the fraction of electrons ionized \cite{Aghetal96}.
Due to density modulation, the kSZ effect has a 
distinct non-Gaussian behavior \cite{GolSpe99}. 
The bulk velocities can be extracted with higher order correlations 
such as the SZ$^2$-kSZ$^2$ statistic of \cite{Coo01a}.
For a given cluster, the temperature fluctuation
is proportional to the line of sight peculiar velocity and when the SZ thermal contribution is removed with frequency information, the kinetic SZ effect should dominate.
In addition to the peculiar motion, any coherent rotation of the cluster will produce a dipole-like temperature distribution toward clusters, especially when the rotational axes is aligned perpendicular to the line of sight \cite{CooChe01}. This allows an extremely useful probe of the cluster rotations and the angular momentum of gas distribution within galaxy clusters.

\vskip 0.2truecm
\begin{quotation}
    \footnotesize\noindent
{\it Acknowledgements:} I  thank Wayne Hu for collaborative work
and organizers of RESCEU 2001 for their hospitality and opportunity to present this work.
This study was supported by the Sherman Fairchild foundation and the DOE.
\end{quotation}

%%%%%%%%%%%%%%%%%%%%%%%%%%%%%%%%%%
%% thebibliography environment %%
%%%%%%%%%%%%%%%%%%%%%%%%%%%%%%%%%

%%%%%%%%%%

\end{document}